\documentclass[12pt]{article}
\usepackage{graphicx}
\usepackage[dvips]{color}
\usepackage{amsmath}
\usepackage{wrapfig}
\begin{document}

\parskip=0.3cm


\vskip 0.5cm \centerline{\bf The Pomeron and Odderon in elastic, inelastic and total cross sections at the LHC} \vskip 0.3cm
\centerline{ L.L.~Jenkovszky$^{a,b}$, A.I.~Lengyel
$^{bc}$, and D.I.~Lontkovskyi$^{d}$}

\vskip 1cm

\centerline{$^a$ \sl BITP, Academy of Sciences of Ukraine, Kyiv
03680 Ukraine} \centerline{$^b$ \sl RMKI, KFKI, 1525 Budapest,
P.O. Box 49, Hungary}
 \centerline{$^{c}$ \sl Institute of Electron
Physics, Nat. Ac. Sc. of Ukraine, Uzhgorod} \centerline{$^d$ \sl
National Taras Shevchenko University of Kyiv, Kyiv, Ukraine}\vskip
0.1cm

\vskip 1cm


\begin{abstract}
A simple model for elastic diffractive hadron scattering,
reproducing the dip-bump structure is used to analyze $pp$ and
$\bar pp$ scattering. The main emphasis is on the delicate and
non-trivial dynamics in the dip-bump region, near $t=-1$ GeV$^2$.
The simplicity of the model and the expected smallness of the
absorption corrections enables one the control of various
contributions to the scattering amplitude, in particular the
interplay between the C-even and C-odd components of the
amplitude, as well as their relative contribution, changing with
$s$ and $t$. The role of the non-linearity of the Regge
trajectories is scrutinized. The ratio of the real to imaginary
parts of the forward amplitude, the ratio of elastic to total
cross sections and the inelastic cross section are calculated.
Predictions for the LHC energy region, where most of the exiting
models will be either confirmed or ruled out, are presented.

\end{abstract}

\vskip 0.1cm




\section{Introduction} \label{s1}
The experimental data on proton-proton elastic and inelastic
scattering emmerging from the measurements at the LHC, call for an
efficient model to fit the data and identify their diffractive
(Pomeron) component \cite{review}. To this end, there is a need
for a reasonably simple and feasible model of the scattering
amplitude, yet satisfying the basic theoretical requirements such
as analyticity, crossing and unitarity. In our opinion, the
expected (dip-bump) structure in the differential cross section is
most critical in discriminating models of high-energy diffraction,
although other observables, such as the rate of the increase of
the total cross sections, the ratio of the elastic to total cross
section, detail concerning the shape of the elastic cross section,
such as its ``break'' at small $|t|$ and flattening at large
$|t|$ are important as well.

We show that, while the contribution from secondary reggeons
is negligible at the LHC, the inclusion of the Odderon is
mandatory, even for the description of $pp$ scattering alone. To
make our analyzis complete, we include in our fits $\bar pp$ data
as well.

Simplicity and efficiency are the main reasons why the model of
Donnachie and Landshoff (DL) \cite{DL} is so popular and useful. A
supercritical Pomeron term, appended with non-leading (secondary)
Reggeon contributions, with linear Regge trajectories describes
elastic scattering data in a wide range of energies at small $-t$.
Due to this simplicity it can be used also as a part of more
complicated inelastic reactions, whenever Regge-factorization
holds. An extension of the DL model and fit can be found in Ref. \cite{KKL}.

Any extension of this model should include:
\begin{itemize}
   \item The dip-bump structure typical to high-energy diffractive processes;
   \item Non-linear Regge trajectories;
   \item Possible Odderon (odd-$C$ asymptotic Regge exchange),and be
   \item Compatible with $s-$ and $t-$ channel unitarity;
\end{itemize}

The first attempt to describe high-energy diffraction, in
particular the appearance of the characteristic dip-bump structure
in the differential cross sections, was made by Chou and Yang
in Ref. \cite{C-Y}, in which the distribution of  matter in the
nuclei was assumed to follow that of the electric charge (form
factors). The original ``geometrical'' Chou and Young model
\cite{C-Y} qualitatively reproduces the $t$ dependence of the
differential cross sections in elastic scattering, however it does
not contain any energy dependence, subsequently introduced by
means of Regge-pole models.

A particularly efficient parametrization of dip was suggested by
 Phillips and Barger in 1973~\cite{PB}, right after its first observation at
the ISR. Their formula reads
\begin{equation}\label{Barger}
\frac{d\sigma}{dt}=|\sqrt A\exp(Bt/2)+\sqrt C\exp(Dt/2+i\phi)|^2,
\end{equation}
where $A,\ B,\ C,\ D$ and $\phi$ are determined independently at
each energy.

 We suggest a simple model that can be used as a handle in studying diffraction
 at the LHC. It combines the simplicity of the above models approach,
 and goes beyond their limitations. Being flexible, it can be modified according to
the experimental needs or theoretical prejudice of its user and
can be considered as the ``minimal model'' of high-energy
scattering while its flexibility gives room for various
generalizations/modifications or further developments (e.g.
unitarization, inclusion of spin degrees of freedom  etc.).

In this paper, we consider the spinless case of the invariant
high-energy scattering amplitude, $A\left(s,t\right)$, where $s$
and $t$ are the usual Mandelstam variables. The basic assumptions
of the model are:

\begin{wraptable}[6]{r}{5.5cm}
\vspace{-24pt}
\begin{center}
    \begin{tabular}{|c|c|c|}
        \hline
      $\alpha(0)\backslash C$& + &-  \\
      \hline
      $>1$& $P$ & $O$ \\
      $<1$& $f$ & $\omega$ \\
      \hline
    \end{tabular}
\end{center}
   \label{tab:sings}\vspace{-14pt}
   \caption{Relative contribution of reggeons to amplitude.}
\end{wraptable}
1. The scattering amplitude is a sum of four terms, two asymptotic
(Pomeron (P) and Odderon (O)), shown in the second row of Table~\ref{tab:sings},
and two non-asymptotic ones or secondary Regge pole contributions
(third row in the same Table).

Viewed vertically, $P$ and $f$ (second column) have positive
$C$-parity, thus entering in the scattering amplitude with the
same sign in $pp$ and $\bar pp$ scattering, while the Odderon and
$\omega$ (third column) have negative $C$-parity, thus entering
$pp$ and $\bar pp$ scattering with opposite signs, as shown below:
    \begin{equation}\label{Eq:Amplitude}
        A\left(s,t\right)_{pp}^{\bar pp}=A_P\left(s,t\right)+
        A_f\left(s,t\right)\pm\left[A_{\omega}\left(s,t\right)+A_O\left(s,t\right)\right],
    \end{equation}
where the symbols $P,\ f,\ O,\  \omega$ stand for the relevant
Regge-pole amplitudes and the super(sub)script, evidently,
indicate $\bar pp (pp)$ scattering with the relevant choice of
the signs in the sum (\ref{Eq:Amplitude}). This sum can be
extended by adding more Reggeons, whose role may become
increasingly important towards lower energies; their contribution
can be effectively absorbed by $f$ and $\omega$ \cite{KKL1}.

 2. We treat the Odderon, the $C$-odd counterpart of the Pomeron on equal
 footing, differing by its $C-$ parity and the values of its parameters
 (to be fitted to the data). We examined also a fit to $pp$ scattering alone, without any Odderon contribution. The (negative) result is presented in Sec.~\ref{sec:fits};

 3. The main subject of our study is the Pomeron, and it is a double pole, or DP \cite{J-Vall, reviews}) lying on a nonlinear trajectory, whose intercept is slightly above one. This choice
is motivated by the unique properties of the DP: it produces
logarithmically rising total cross sections at unit Pomeron
intercept. By letting $\alpha_P\left(0\right)>1,$ we allow for a
faster rise of the total cross section \footnote{A supercritical
Pomeron trajectory,  $\alpha_P(0)>1$ in the DP is required by the
observed rise of the ratio $\sigma_{el}/\sigma_{tot},$ or,
equivalently, departure form geometrical scaling \cite{VJS}.},
although the intercept is about half that in the DL model since the
double pole (or dipole) itself drives the rise in energy. Due to its geometric form
(see below) the DP reproduces itself against unitarity (eikonal)
corrections. As a consequence, these corrections are small, and
one can use the model at the ``Born level'' without complicated
(and ambiguous) unitarity (rescattering) corrections. DP combines
the properties of Regge poles and of the geometric approach,
initiated by Chou and Yang, see \cite{C-Y}.

Higher order (triple) multipoles and their interpretation as a finite-rang ladder diagram can be found in Ref. \cite{KLT}.

4. Regge trajectories are non-linear complex functions. In a
limited range and with limited precision, they can be approximated by
 linear trajectories (which is a common practice, reasonable when non-linear effects can be neglected). This nonlinearity is manifest e.g. as the ``break'' i.e. a change the slope $\Delta B \approx 2$ GeV$^2$ around $t\approx-0.1$ GeV$^2$  and at large $|t|$, beyond the second maximum,  $|t|>2$ GeV$^2$, where the cross section flattens and the trajectories are expected to slowdown logarithmically.


A simple mechanism of the diffractive dip-bump structure combining
geometrical features and Regge behavior was suggested in Ref.
\cite{J-Vall}. In that model, the dip is generated by the Pomeron
contribution. The relevant Pomeron is a double pole 
arises from the interference between this dipole with a simple
one, it is accompanied by. The dip-bump in the model shows correct
dynamics, that is it develops from a shoulder, progressively
deepening in the ISR energy region. As energy increases further, the
dip is filled by the Odderon contribution. At low energies the
contribution from non-leading, ``secondary'' Reggeons is also
present.

Physically, the components of the Pomeron have the following
interpretation: the first term in Eq. (\ref{GP}) is a Gaussian in
the impact parameter representation, while the second term
contains absorption corrections generating the dip.

    The dipole Pomeron produces logarithmically rising total cross
sections and nearly constant ratio of $\sigma_{el}/\sigma_{tot}$ at
unit Pomeron intercept, $\alpha_P\left(0\right)=1.$ While a mild,
logarithmic increase of $\sigma_{tot}$ does not contradict the data,
the rise of the ratio $\sigma_{el}/\sigma_{tot}$ beyond the SPS
energies requires a supercritical DP intercept,
$\alpha_P\left(0\right)=1+\delta,$ where $\delta$ is a small
parameter $\alpha_P(0)\approx 0.05$. Thus DP is about ``twice softer''
then that of Donnachie-Landshoff \cite{DL}, in which
$\alpha_P(0)\approx 0.08.$

In spite of a great varieties of models for high-energy
diffraction (for a recent review see \cite{review}), only a few of
them attempted to attack the complicated and delicate mechanism of
the diffraction structure. In the 80-ies and early 90-ies, DP was
fitted to the ISR, SPS and Tevatron data, see \cite{2, 3, KLT}
and \cite{VJS} for earlier references. Now we find it appropriate
to revise the state of the art in this field, to update the
earlier fits,  analyze the ongoing measurements at the LHC and/or
make further predictions. We revise the existing estimates of the
Pomeron contribution to the cross sections as a functions of $s$
and $t$ and argue that while the contribution from non-leading
trajectories in the nearly forward region is negligible
(smaller than the experimental uncertainties), the Odderon may be
important, especially in the non-forward direction.

\section{The model} \label{s1}
\vskip 0.2cm We use the normalization:
\begin{equation}\label{norm}
{d\sigma\over{dt}}={\pi\over s^2}|A(s,t)|^2\ \  {\rm and}\ \
\sigma_{tot}={4\pi\over s}\Im m A(s,t)\Bigl.\Bigr|_{t=0}\ .
\end{equation}
Neglecting spin dependence, the invariant
proton(antiproton)-proton elastic scattering amplitude is that of Eq. (\ref{Eq:Amplitude}).
The secondary Reggeons are
parametrized in a standard way \cite{KKL, KKL1}, with linear Regge trajectories and
exponential residua, where $R$ denotes $f$ or $\omega$ - the
principal non-leading contributions to $pp$ or $\bar p p$
scattering:
\begin{equation}\label{Reggeons}
A_R\left(s,t\right)=a_R{\rm e}^{-i\pi\alpha_R\left(t\right)/2}{\rm e}
^{b_Rt}\Bigl(s/s_0\Bigr)^{\alpha_R\left(t\right)},
\end{equation}
with $\alpha_f\left(t\right)=0.70+0.84t$ and
$\alpha_{\omega}\left(t\right)=0.43+0.93t;$ the values of other
parameters of the Reggeons are quoted in Tables~\ref{tab:fitParam1},~\ref{tab:fitParam2},~\ref{tab:fitParam3}.

As argued in the Introduction, the Pomeron is a dipole in the $j-$plane
\begin{equation}\label{Pomeron}
A_P(s,t)={d\over{d\alpha_P}}\Bigl[{\rm
e}^{-i\pi\alpha_P/2}G(\alpha_P)\Bigl(s/s_0\Bigr)^{\alpha_P}\Bigr]=
\end{equation}
$${\rm
e}^{-i\pi\alpha_P(t)/2}\Bigl(s/s_0\Bigr)^{\alpha_P(t)}\Bigl[G'(\alpha_P)+\Bigl(L-i\pi
/2\Bigr)G(\alpha_P)\Bigr].$$
Since the first term in squared brackets determines the shape of
the cone, one fixes
\begin{equation} \label{residue} G'(\alpha_P)=-a_P{\rm
e}^{b_P[\alpha_P-1]},\end{equation} where $G(\alpha_P)$ is recovered
by integration, and, as a consequence, the Pomeron amplitude Eq.
(\ref{Pomeron}) can be rewritten in the following ``geometrical''
form (for the details of the
calculations see \cite{VJS} and references therein)
\begin{equation}\label{GP}
A_P(s,t)=i{a_P\ s\over{b_P\ s_0}}[r_1^2(s){\rm
e}^{r^2_1(s)[\alpha_P-1]}-\varepsilon_P r_2^2(s){\rm
e}^{r^2_2(s)[\alpha_P-1]}],
\end{equation} where
$r_1^2(s)=b_P+L-i\pi/2,\ \ r_2^2(s)=L-i\pi/2,\ \ L\equiv
\ln(s/s_0).$

The main features of the nonlinear trajectories are: 1) presence of a threshold singularity required by $t-$channel unitarity and responsible for the change of the slope in the exponential cone (the so-called ``break'') near $t=-0.1$ GeV$^2$ \cite{Cohen}, and 2) logarithmic asymptotic behavior  providing for a power fall-off of the cross sections in the ``hard'' region \cite{Zurab}. The combination of theses properties is however not unique, see \cite{reviews}.

We examine representative examples of the Pomeron trajectories, namely: 1) Linear Eq.~(\ref{eq:tr1}); 2) With a square-root threshold, Eq.~(\ref{eq:tr2}), required by $t-$channel unitarity and accounting for the small-$t$ ``break'' \cite{Cohen}, as well as the possible ``Orear'', $~e^{\sqrt{-t}}$ behavior in the second cone; and 3) A logarithmic one, Eq.~(\ref{eq:tr3}) anticipating possible ``hard effects'' at large $|t|$ (in fact, our fits (see below) do not show the expected large-$t$ logarithmic regime in the transition region $|t|<8$ GeV$^2$.
\begin{align}
\alpha_P\equiv \alpha_P(t) &=
1+\delta_P+\alpha_{1P}t,\tag{TR.1}\label{eq:tr1}
\\
\alpha_P\equiv \alpha_P(t) &=
1+\delta_P+\alpha_{1P}t - \alpha_{2P}\left(\sqrt{4\alpha_{3P}^2-t}-2\alpha_{3P}\right),\tag{TR.2}\label{eq:tr2}
\\
\alpha_P\equiv \alpha_P(t) &=
1+\delta_P-\alpha_{1P}\ln\left(1-\alpha_{2P}t\right).\tag{TR.3}\label{eq:tr3}
\end{align}
Alternatives choices for the nonlinear trajectories and fits can be found e.g. in \cite{DLM}.


An important property of the DP Eq.~(\ref{GP}) is the presence of
absorptions, quantified by the value of the parameter
$\varepsilon_P$ in Eq.~(\ref{GP}); this property, together with the non-linear nature of the trajectories, justifies the neglect of the rescattering corrections.  More details can be found e.g.
in Ref. \cite{VJS}.)

The unknown Odderon contribution is assumed to be of the same form
as that of the Pomeron, Eqs.~(\ref{Pomeron}),~(\ref{GP}), apart from different values of adjustable parameters (labeled by the subscript ``$O$''). Also only one trajectory of type~(\ref{eq:tr1}) is considered for the Odderon.
\begin{equation}\label{Odd}
A_O(s,t)={a_O\ s\over{b_O\ s_0}}[r_{1O}^2(s){\rm
e}^{r^2_{1O}(s)[\alpha_O-1]}],
\end{equation}
The adjustable parameters are: $\delta_P,\ \alpha_{iP},\ 
a_P,\ b_P,\ \varepsilon_P$ for the Pomeron and
$\delta_O,\ \alpha_{iO},\ a_O,\ b_O$ for
the Odderon. The results of the fitting procedure is  presented
below.

\section{Fitting strategy}
The model contains from 14 to 16 parameters (depending on the choice of the trajectories) to be fitted to 1024 data points simultaneously in $s$ and $t$.
By a straightforward minimization one has little chances to find the solution, because of possible correlations between different contribution and the parameters, including the $P-f$ and $O-\omega$ mixing and the unbalanced role of different contributions/data points. 
Although we apply the best global fit (minimal $\chi^2$) as a formal criterion for the valid description, we are primarily interested in the dip region, critical for the identification of the  Pomeron and the Odderon. As mentioned in the Introduction, we perform also a fit to $pp$ data alone, see Subsection 4.1, to see whether the observed dynamics of dip can be reproduced by the Pomeron alone. The contribution to the global $\chi^2$ from tiny effects, such as the small-$|t|$ ``break'' in the first (and second) cone, possible oscillations in the slope of the cone(s) etc. should not corrupt the study of the dynamics in the dip-bump region.


The following kinematical regions and relevant datasets were involved in the fitting procedure:
23, 32, 45, 53, 62 GeV for $pp$ scattering  \cite{Amaldi:1979kd,Albrow:1976sv,Breakstone:1984te,Breakstone:1985pe}
and 31, 53, 62, 546, 630, 1800 GeV for $\bar{p}p$ scattering
\cite{Bozzo:1985th,Bernard:1986ye,Amos:1990fw,Abe:1993xx}. These datasets were compiled in a single one in~\cite{data}. The differential elastic scattering cross sections were further constrained to cover momentum transfer range $|t|=$0.1 --- 8 GeV$^2$.

To avoid false $\chi^2$ minima, we proceed step-by-step. We start with a fit to the to the forward data: the total cross section and the ratio $\rho=\Re e A(s,t=0)/\Im m A(s,t=0)$ with the Pomeron contribution alone, by assuming that the contribution from the Odderon is small and no absorption in the Pomeron amplitude, $\varepsilon_P=0$. The forward data are sensitive only to the parameters like $a_P, \delta_p$, therefore we fit them at first. Using obtained values of the parameters as an initial point we proceed with the fitting of the $pp$ and $\bar pp$ differential cross sections data in the first cone $|t| < 0.5$~GeV$^2$, thus applying further constraints on previously mentioned parameters, $b_P$ and $\alpha_{iP}$. These fits give satisfactory description ($\chi^2/NDF\approx 1.5$) of the total cross sections, ratios of real to imaginary part of the forward amplitude and of the first cone in both $pp$ and $\bar pp$ cases for each energy. To describe the second cone and the dip-bump structure, we fit the $\varepsilon_P$ and the Pomeron's trajectory parameters: $\alpha_{iP}$. Next we assume that a shelf, which is clearly seen in $p\bar p$ data at 546 and 630~GeV, is generated by the Odderon. Since there is no information about Odderon's structure we fit all its parameters simultaneously, but fixing the Pomeron. After these steps to polish out the minimum we release all parameters of the primary reggeons and add the secondary regeons for the final fit.

To find the best set of parameters we minimize a combined $\chi^2=\chi^2_{tot}+\chi^2_{\rho}+\chi^2_{pp}+\chi^2_{p\bar p}$ using the MINUIT \cite{MINUIT} code.
The obtained minimal value of $\chi^2$ for the model with trajectory~(\ref{eq:tr1}) corresponds to $\chi^2/NDF$ = 3. Details of the fit results for different trajectories are summarized in  Tables~(\ref{tab:fitParam1},\ref{tab:fitParam2},\ref{tab:fitParam3}).

We performed an error analysis on estimated values of the model parameters, namely
we propagated the experimental uncertainties of the measured quantities into uncertainties
of the parameters.
Using the estimated parameters errors we propagated them
into an uncertainty on predicted cross sections at the LHC energies. We sample a sufficiently large
number of predicted cross sections with different values of model parameters drawn from a
gaussian distribution with mean values equal to the nominal values of the parameters values and variances equal to
their uncertainties.

Finally we note that a the best fit to the data does not necessarily
implies the best physical model. For example, the includion of spin \cite{TT} may affect any seemingly perfect fit to the data. In our opinion, such a minimization procedure improves our understanding of the
physical meaning of each term introduced phenomenologically in the amplitude.

\section{Results}\label{sec:fits}
\subsection{Fits without the Odderon}
\begin{figure}[htbp!]
\center{
        \includegraphics[width=0.45\linewidth]{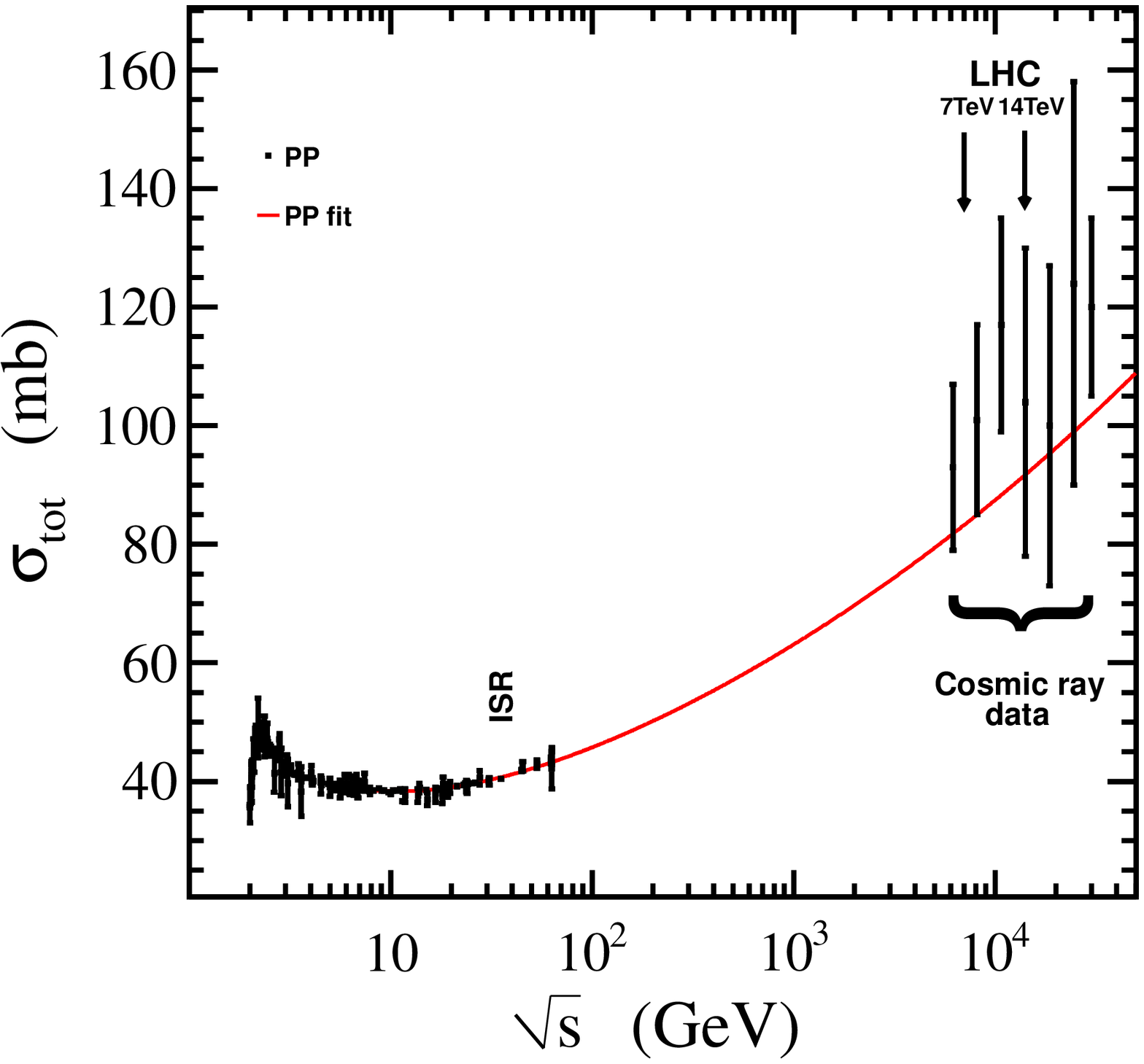}
        \includegraphics[width=0.45\linewidth]{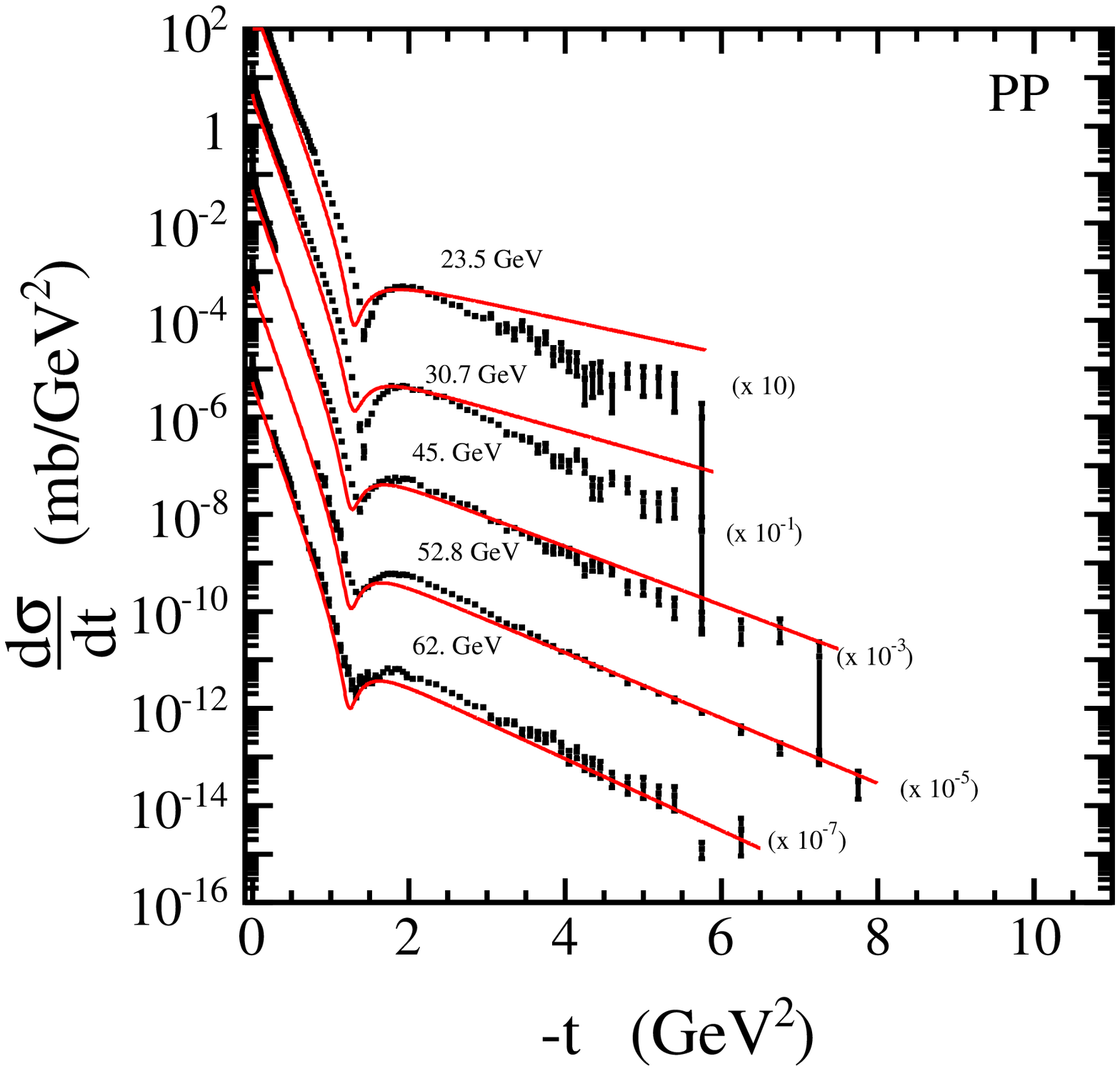}\\
        (a)\hspace{0.45\linewidth}(b)
}
\caption{
(a) Total $pp$ cross section calculated in the model, Eqs. (2-8,~\ref{eq:tr1}), without the Odderon term and fitted to the data in the range $\sqrt{s}$ = 5 --- 30~TeV;
(b) Differential $pp$ cross sections calculated in model, Eqs. (2-8,~\ref{eq:tr1}), without the Odderon term and fitted to the data in the range $-t$ = 0.1 --- 8~GeV$^2$.
\label{fig:totpp}
}
\end{figure}

To check the role of the Odderon, we first fit only $pp$ scattering without any Odderon (that is supposed to fill the dip in $\bar pp$). The best fit is shown in Figs~\ref{fig:totpp}~(a,b), demonstrating that, while the Pomeron appended with sub-leading reggeons reproduces the dip for several energies, namely 45, 53, 62~GeV, it fails otherwise (we remind that the deepening of dip is not monotonic: after the minimum at $\sqrt s \approx 35$ GeV the trend gets reversed). The presence of the Odderon seems inevitable. Henceforth we use the complete amplitude Eq.~(\ref{Eq:Amplitude}), including the Odderon.
\subsection{Elastic cross sections and dip-bump at the LHC}
\begin{figure}[htbp!]
\vspace{-30pt}
\center{
        \includegraphics[width=0.45\linewidth]{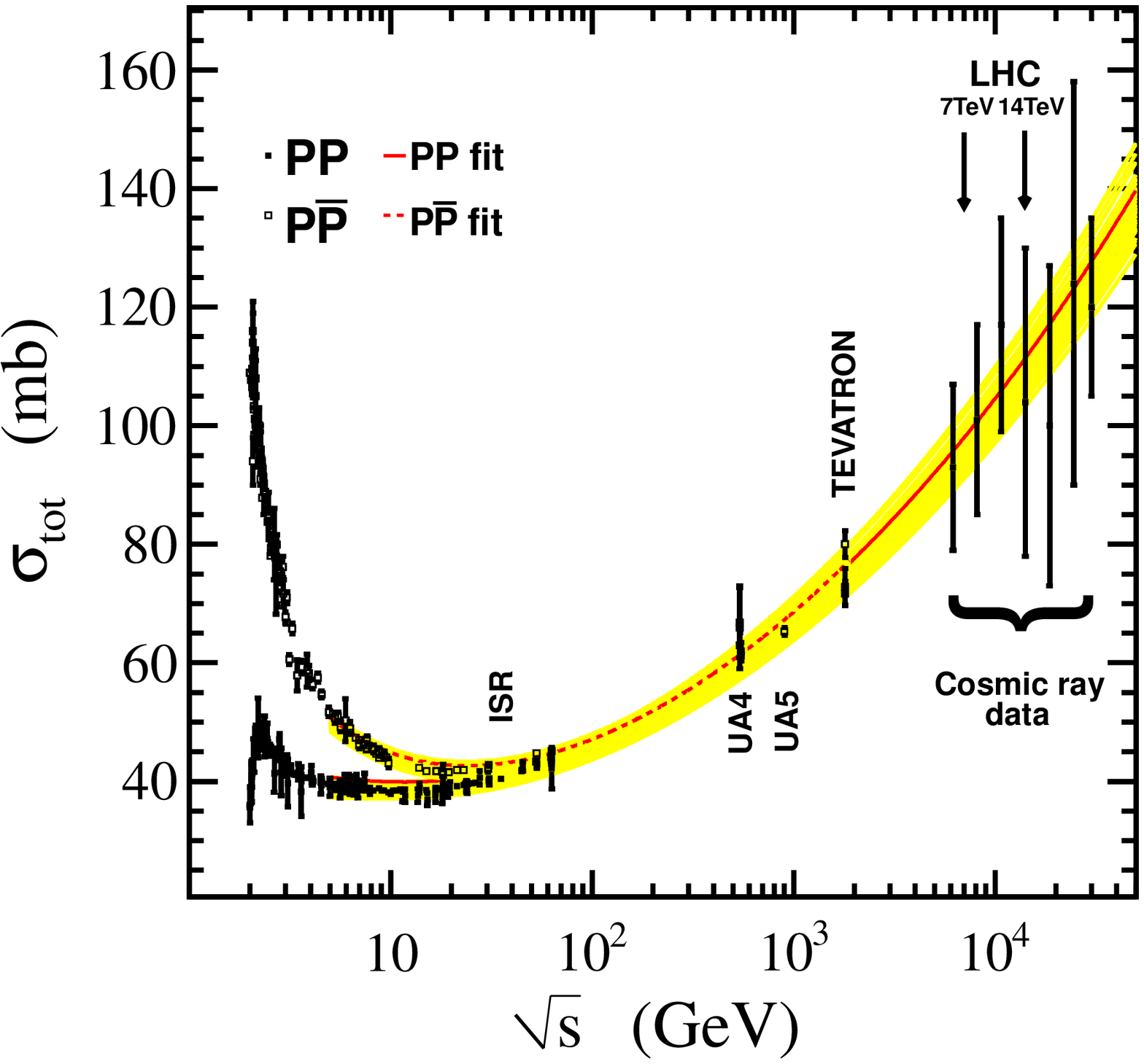}
        \includegraphics[width=0.45\linewidth]{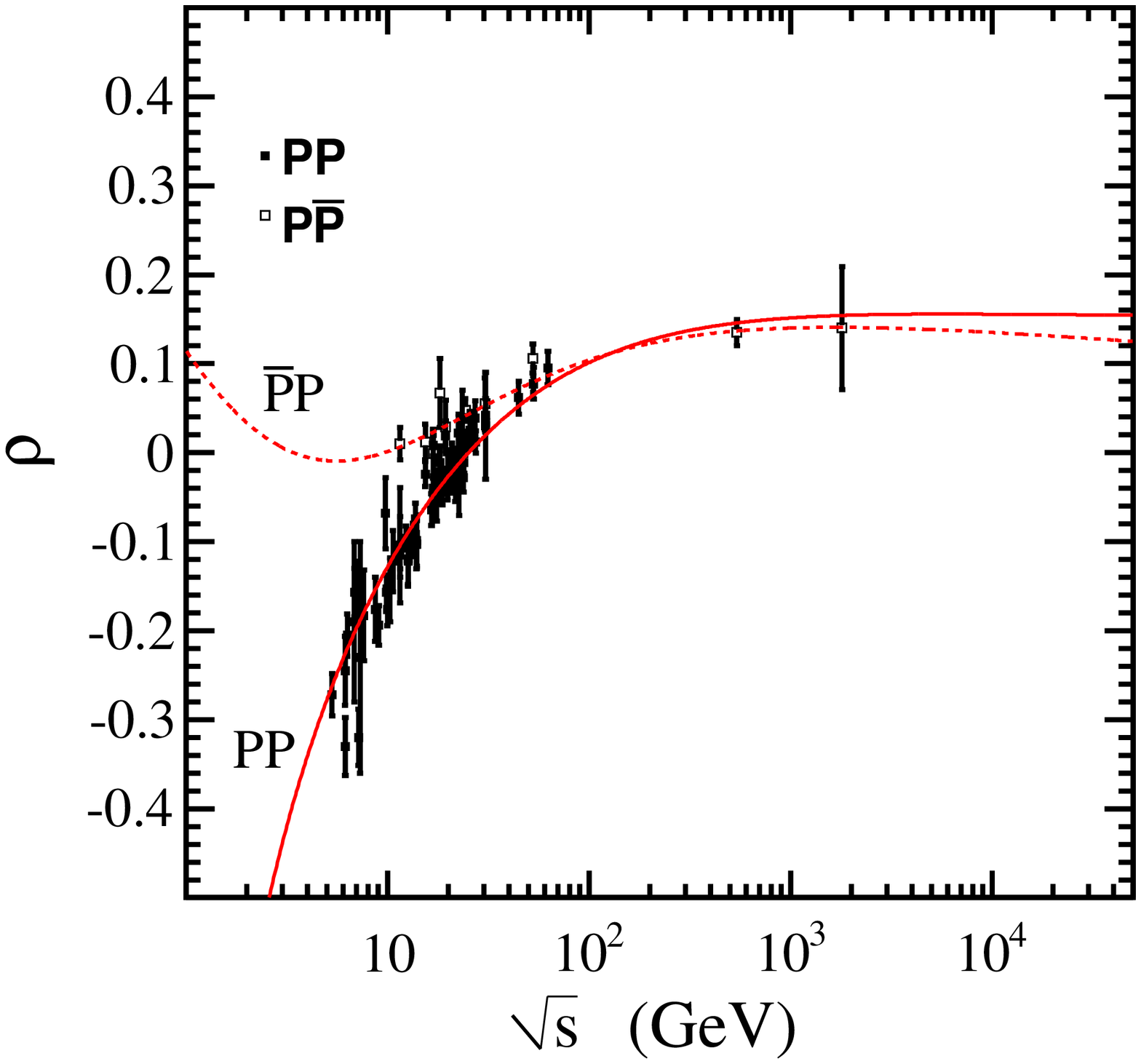}\\
        (a)\hspace{0.45\linewidth}(b)
}
\caption{
(a) Total $pp$ and $\bar{p}p$ cross sections calculated in model, Eqs. (2-8,~\ref{eq:tr1}), and fitted to the data in the range $\sqrt{s}$ = 5 --- 30~TeV and 5 GeV --- 1.8~TeV, respectively.
(b) Ratio of the real to imaginary part of the forward amplitude for $pp$ and $\bar{p}p$, calculated in model and fitted to the data. The curves correspond to calculations with the parameters shown in Table~(\ref{tab:fitParam1}).}
\label{fig:rho_all}
\end{figure}

\begin{figure}[htbp!]
\vspace{-30pt}
\center{
        \includegraphics[width=0.45\linewidth]{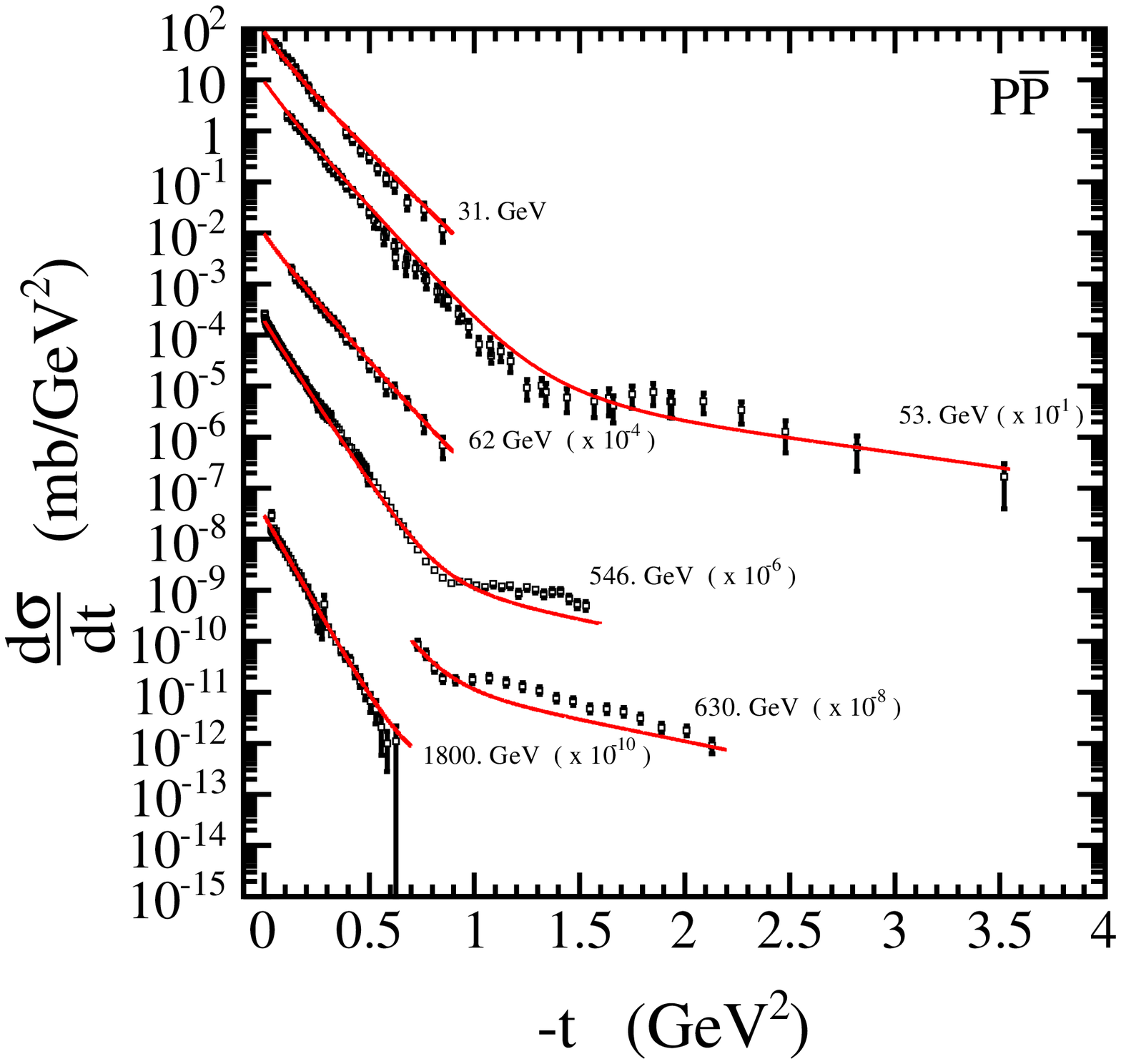}
        \includegraphics[width=0.45\linewidth]{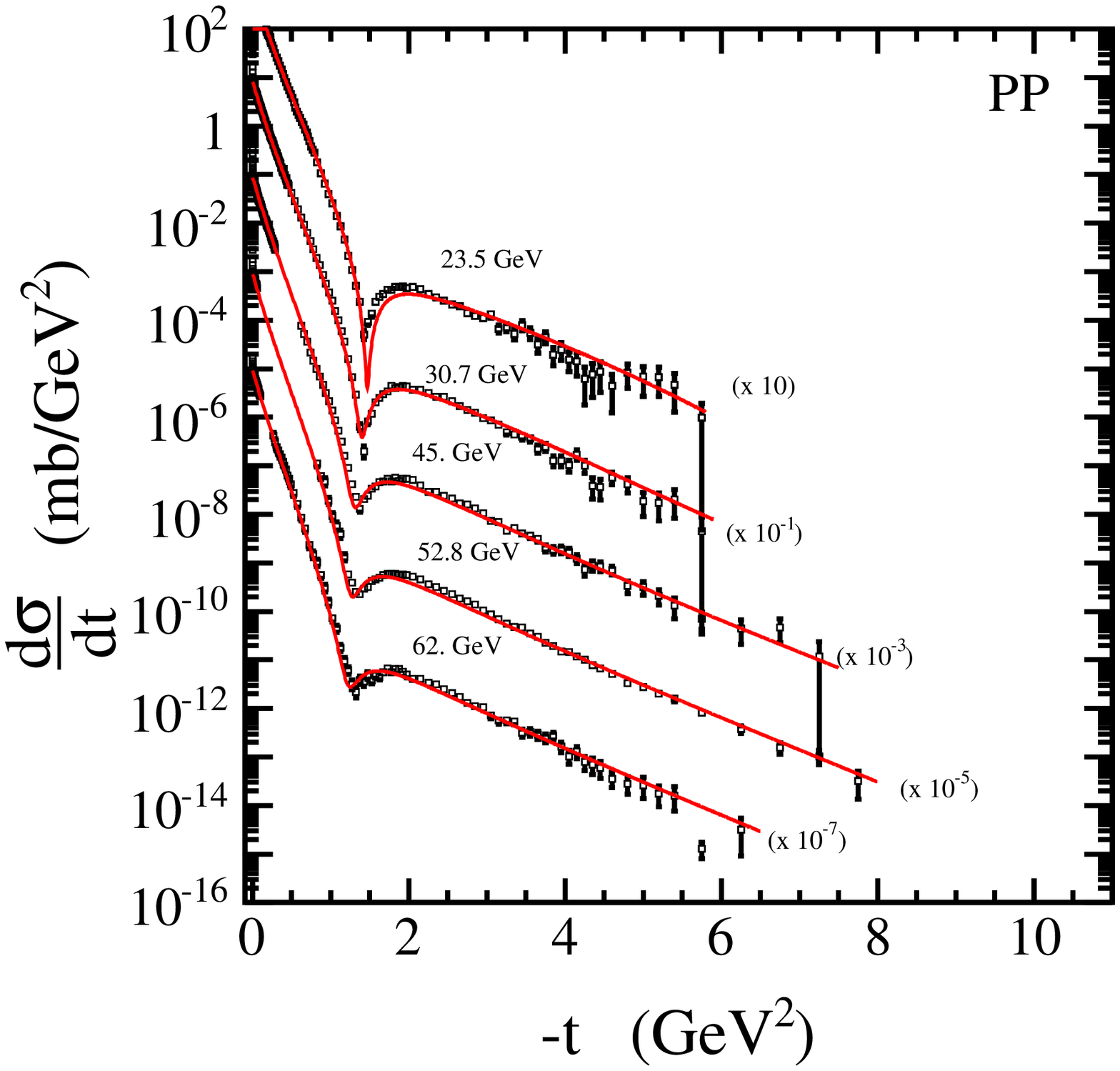}\\
                (a)\hspace{0.45\linewidth}(b)
}
\caption{
(a) $\bar{p}p$ differential cross sections calculated
in model, Eqs. (2-8,~\ref{eq:tr1}), and fitted to the data, and fitted to the data in the range $-t$ = 0.1 --- 8~GeV$^2$.
(b) $pp$ differential cross sections calculated
in the model and fitted to the data. The curves present calculations with the parameters shown in Table (\ref{tab:fitParam1}).
}
\label{fig:pp_all}
\end{figure}

\begin{figure}[htbp!]
\vspace{-30pt}
\center{ \includegraphics[angle=0,width=0.35\textwidth]{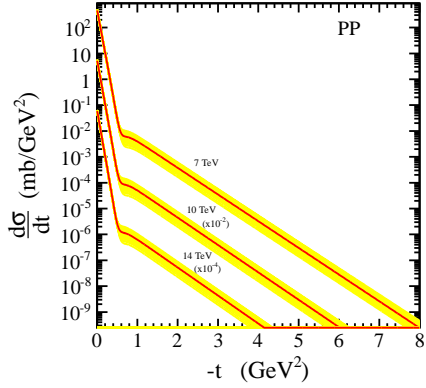}}
\caption{Predictions for the $pp$ differential cross section calculated
in model, Eqs. (2-8,~\ref{eq:tr1}) for three different LHC energies. Curves present calculations with the parameters shown in Table (\ref{tab:fitParam1}). The width of the yellow band corresponds to the uncertainty in the cross sections, estimated as described in Sec. 3.}
\label{fig:LHC}
\end{figure}

Figure~\ref{fig:rho_all}~(a) shows the $pp$ and $\bar{p}p$ total elastic scattering cross section calculated in model with the parameters presented in Table~\ref{tab:fitParam1}. On this plot yellow band represents statistical uncertainties on the calculated values of the total cross section. Figure~\ref{fig:rho_all}~(b) shows the ratio of the real to imaginary part of the forward amplitude. The model with a linear trajectory sufficiently well describes the forward quantities in a wide range of collision energies for $pp$ and $\bar{p}p$. Different choices of the Pomeron trajectory give similar description of the data. The values of parameters fitted with different trajectory forms are summarized in Tables~\ref{tab:fitParam1},\ref{tab:fitParam2},\ref{tab:fitParam3}. Figures~\ref{fig:pp_all}~(a,b) show the fitted $\bar{p}p$ and $pp$ differential elastic scattering cross sections. The model reasonably describe both reactions with slight excess around the dip region at $\sqrt{s}=$23~GeV for $pp$ scattering and small deviations for $|t|>$ 1~GeV$^2$ in $\bar{p}p$. In Figure~\ref{fig:LHC} predictions for three different center of mass energies are shown. The yellow area exhibits the statistical uncertainty on the calculations, described earlier. Calculations are characterized by an approximately exponential fall-off in range $0 < |t| < 8$ GeV$^2$, with the slope change around $-t\approx$ 0.6~GeV$^2$. The dip moved towards lower momentum transfer and became almost filled by the Odderon contribution. Predictions on elastic scattering at the LHC are summarized in the Table~\ref{tab:LHCpredictions}.
\subsection{Inelastic cross section $\sigma_{in}(s)$ and the ratio $\sigma_{el}/\sigma_{tot}$}
\begin{figure}[htbp!]
\vspace{-10pt}
\center{
\includegraphics[width=0.4\linewidth]{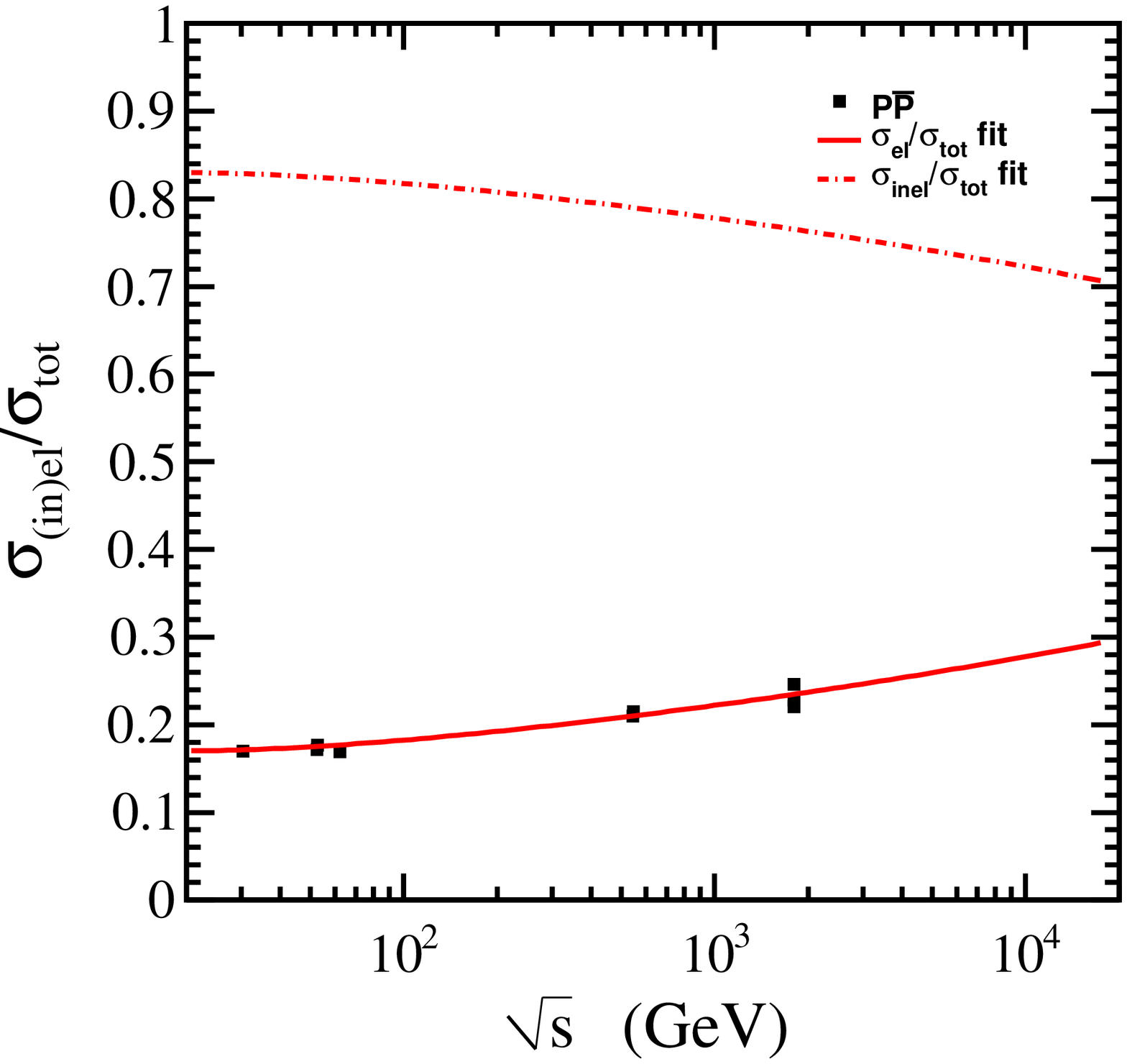}
\includegraphics[width=0.4\linewidth]{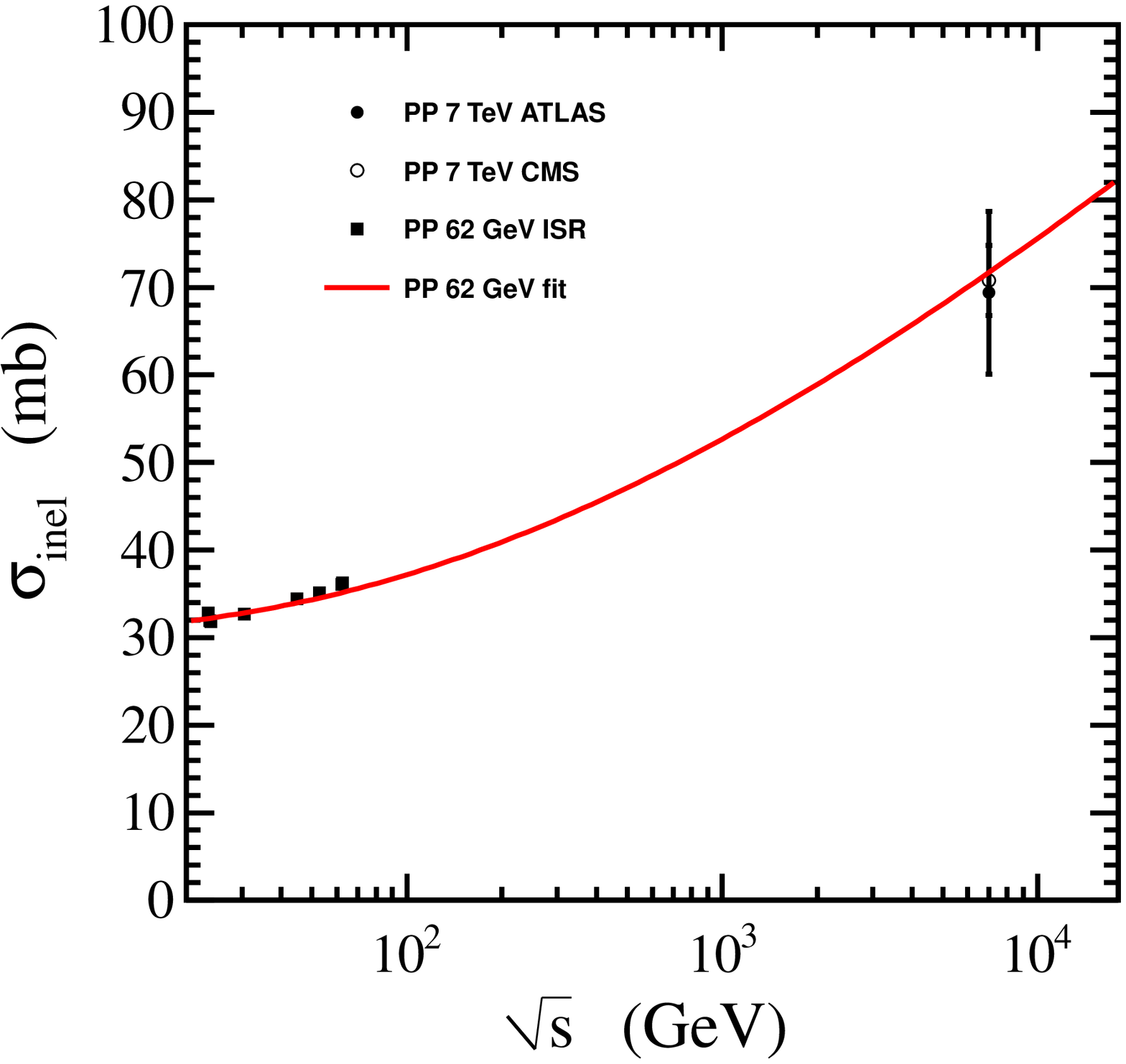}\\
       (a)\hspace{0.45\linewidth}(b)
}
\caption{(a) The ratios $\frac{\sigma_{el}}{\sigma_{tot}}$ and $\frac{\sigma_{inel}}{\sigma_{tot}}$ calculated in model using the trajectory (\ref{eq:tr1}). (b) Predictions for the $pp$ inelastic cross section calculated in model, Eq. (\ref{norm}). The curves correspond to the calculations with the parameters quoted in Table~(\ref{tab:fitParam1}). 
}
\label{fig:inel}
\end{figure}

We calculate  $\sigma_{el}(s)$ by  integration
\begin{equation}\label{eq:el}
\sigma_{el}=\int_{t_{min}}^{t_{max}}(d\sigma/dt\, dt),
\end{equation}
where formally $t_{min}=-s/2$ and $t_{max}=t_{threshold}$. Since the integral is saturated basically by the first cone, we use $t_{max}=0$ and  $t_{min}=-25$ GeV$^2$ ($t_{min}=-3$ GeV$^2$ would do as well.) Next we calculate $\sigma_{in}(s)=\sigma_{tot}-\sigma_{el}$ The calculated ratios $\sigma_{el}(s)/\sigma_{tot}(s)$ and $\sigma_{in}(s)/\sigma_{tot}(s)$ are shown in Figure~\ref{fig:inel} (a). Figure~\ref{fig:inel} (b) shows $pp$ inelastic cross section. On that figure recent measurements by ATLAS \cite{ATLAS} and CMS \cite{CMS} are also shown. The model is found to be in a good agreement with $\bar{p}p$ and low energy $pp$ data as well as with the the newest measurements at 7 TeV (not fitted).

\newpage
\subsection{Local Nuclear Slope}\label{loc_sl}
Having fitted the parameters to the data on differential and total cross sections as well as on the ratio $\rho$, we proceed to calculate the local slope


\begin{figure}[tbh!]
\center{ \includegraphics[width=0.45\textwidth]{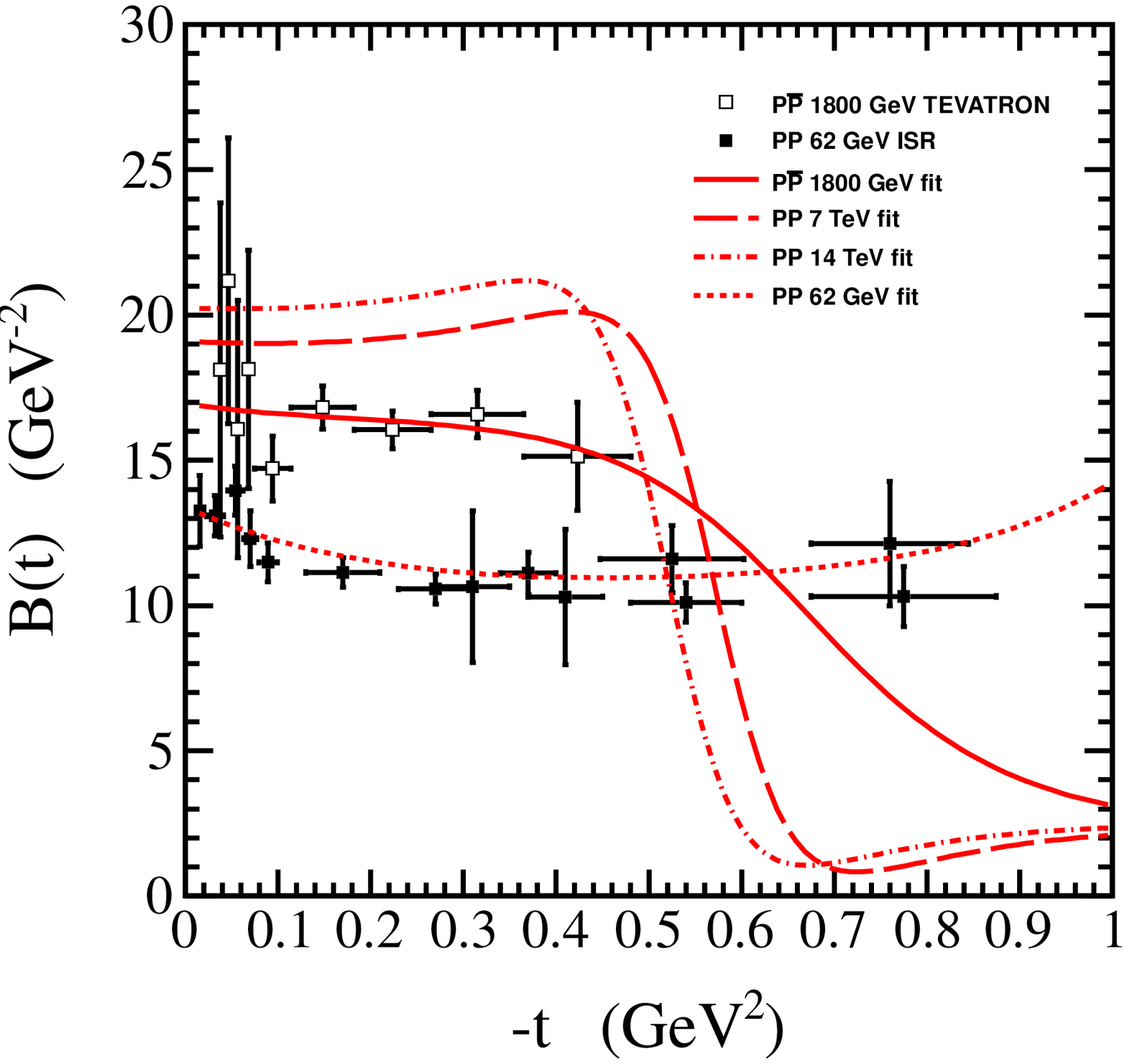}\includegraphics[width=0.45\textwidth]{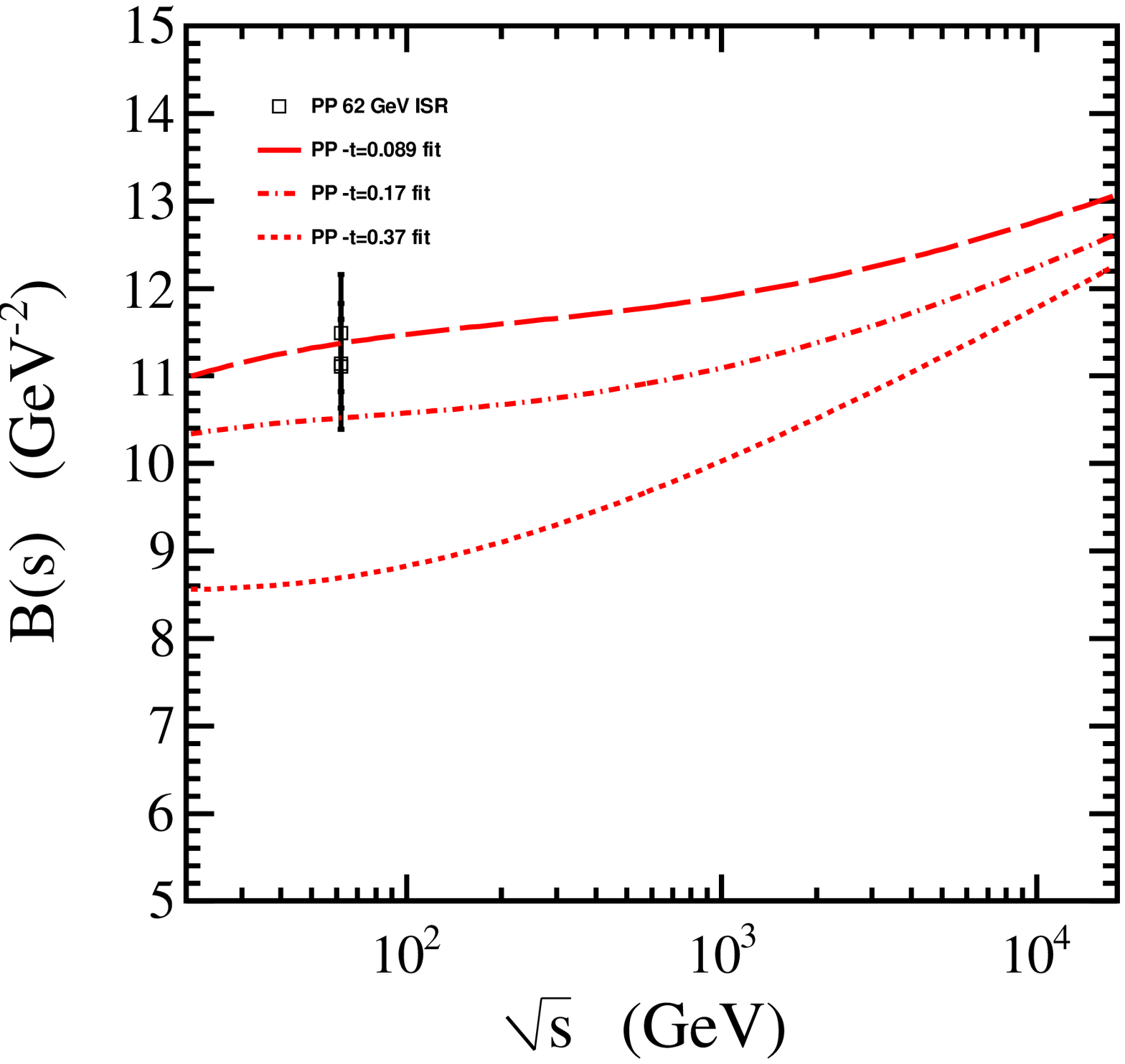}}
\caption{$pp$ and $p\bar p$ slope $B(s,t)$ calculated from the
model, Eq. (\ref{Pomeron}).} \label{fig:slope}
\end{figure}

\begin{equation}\label{Nuclslope}
B(s,t)={d\over{dt}}\left(\ln {d\sigma(s,t)\over{dt}}\right).
\end{equation}
It is a sensitive tool to investigate the fine structure of the
cone.

The purpose of the present calculations of $B(s,t)$ is to
reproduce and predict the behavior of the slope (usually not
fitted) at different energies, including those of the LHC.
We have
calculated the local nuclear slope $B(s,t)$ within the present model
using the parameters from Table~\ref{tab:fitParam1}, and compared it with the
``experimental'' local nuclear slope, obtained by an ``overlapping
bins'' procedure \cite{kontros}. To calculate $B(s,t)$, we use the
approximate formula
\begin{equation}\label{slope}
B(s,t)= {{1\over{2\Delta t}}{({d\sigma\over{dt}}(s,t+\Delta t) -
{d\sigma\over{dt}}(s,t-\Delta
t))}\over{{d\sigma\over{dt}}(s,t)}}.
\end{equation}

The results of the calculations are shown in Fig.~\ref{fig:slope}. One can see that
$B(s,t)$ agrees with the experimental data for 63 GeV pp and 1800 GeV $\bar{}pp$. With
increasing energy, the  curvature decreases and changes the sign
when the energy exceeds $\sim 2$ TeV.

\subsection{Pomeron dominance at the LHC}\label{pom_dom}
A basic problem in studying the Pomeron is its identification i.e.
its discrimination from other contributions.
 We try to answer the important question: where
 (in $s$ and in $t$) and to what extent are the elastic data from the LHC dominated by the Pomeron contribution? The answer to this question
is of practical importance since, by Regge-factorization, it can
be used in other diffractive processes, such as diffraction
dissociation. It is also of conceptual interest in our definition
and understanding of the phenomenon of high-energy diffraction. An
earlier attempt to answer this question was made in
Ref.\cite{Ilyin}.

\begin{figure} [tbh!]
\begin{center}
	\includegraphics[width=0.45\textwidth]{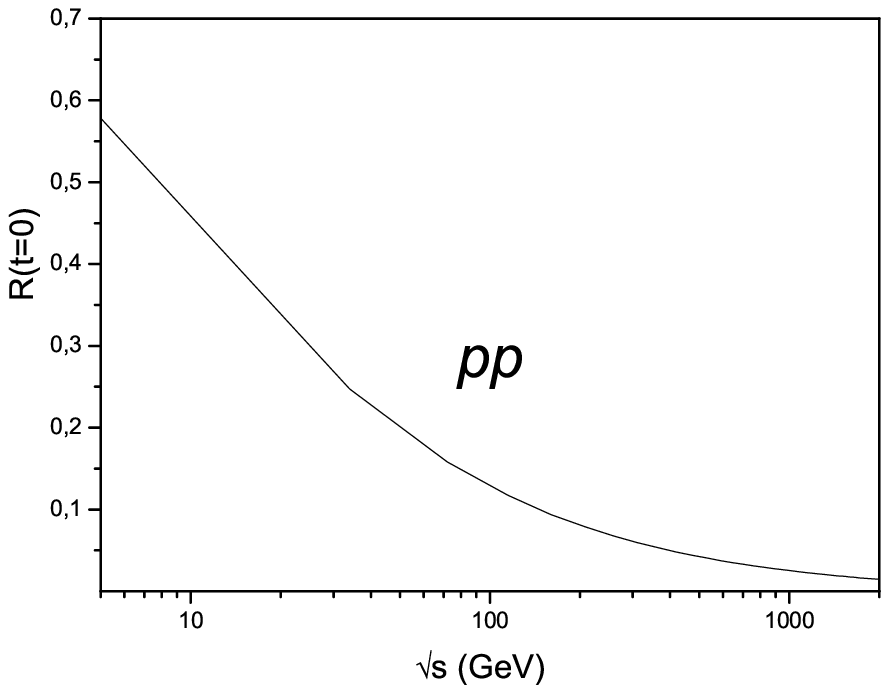}
	\includegraphics[width=0.45\textwidth]{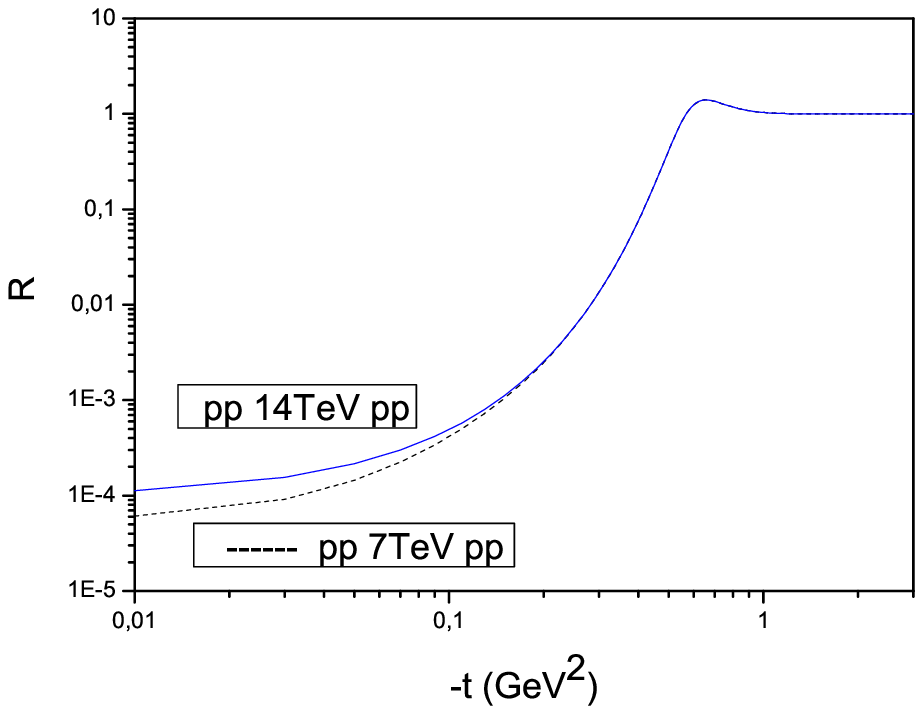}\\
	(a)\hspace{0.45\linewidth}(b)
\end{center}
\caption {(a) Relative importance of the non-leading (non-Pomeron)
contributions R(s,t=0)  to the pp total cross-sections versus
energy. (b) Relative importance of non-leading (non-Pomeron)
contribution R(s,t) to the pp differential cross-sections
calculated versus $t$.}
\label{fig:Rt0}
\end{figure}


First we show the energy variation of the relative importance
of the Pomeron with respect to contributions from the secondary
trajectories and the Odderon. In the case of the $pp$ total
cross-section, we calculated the ratio:

\begin{equation}\label{relative pomeron}
R(s,t=0)={{\Im m (A(s,t)-A_P(s,t))\over{\Im A(s,t)}}},
\end{equation}
where the total scattering amplitude ${A}$ includes the Pomeron
contribution ${A_P}$ plus the contribution from the secondary
Reggeons and the Odderon. The results are shown in Fig.~\ref{fig:Rt0}~(a).

We conclude that starting from the Tevatron energy region, the
relative contribution of the non-Pomeron terms to the total
cross-section becomes smaller than the experimental uncertainty and
hence at higher energies they may be completely neglected,
irrespective of the model used. Such a discrimination (between
Pomeron and non-Pomeron contributions) is more problematic in the
non-forward direction, where the real and imaginary parts of
various components of the scattering amplitude behave in a
different way and the phase can not be controlled experimentally.
Similarly, we calculate the ratio for non-forward scattering ${(t
\neq 1)}$:

\begin{equation}\label{relative pomeron t non forward}
R(s,t)={{\left|(A(s,t)-A_P(s,t)\right|^2\over{\left|A(s,t)\right|^2}}}.
\end{equation}

We have calculated this ratio for $pp$ scattering at LHC energies
within the framework of the model. The results are shown in Fig.~\ref{fig:Rt0}~(b),
where $R(s,t)$ is plotted versus $|t| < $1 GeV$^2$ at the energy equal to 14 TeV.
The common feature of these results is that the Reggeons and the
Odderon contributions increase in the vicinity of the dip
(shoulder in the case of $pp$ scattering).

\newpage
\section{Conclusions}
The aim of the present paper was the identification of the Pomeron contribution at the LHC. This was possible due to the simplicity of the model, which has the important property of reproducing itself (approximately) against unitarity (absorption) corrections (for more details see
\cite{J-Vall, 2,3} and references therein).

We have presented the ``minimal version'' of the DP model. It can be
further extended, refined and improved, its basic features remaining intact. It should be remembered however that any detailed study of diffraction should include also the poorly known spin effects \cite{TT}. They  small in the forward direction but may  increase away from $t=0,$ thus affecting the details of any fit to the data.

Further studies of the small-$t$ curvature (the ``break'' or fine
structure of the Pomeron), with the Coulombic term added will
reproduce (and predict) the behavior of  elastic cross sections in the Coulomb interference region, while the intermediate- and large-$t$ behavior can be accounted for by using a Pomeron trajectory with a logarithmic asymptotics.

We conclude that:

1. A single shallow dip (in fact, a break) is expected in the elastic differential cross section at the LHC, followed by a smooth behavior in $t$.

2. The Odderon is indispensable in the description of elastic scattering. Its relative contribution, small in the forward direction, increases away from $t=0$, becoming  particularly important in the dip-bump region.

3. The contribution from the non-leading (secondary) Regge trajectories can be neglected in the kinematic region of the LHC measurements. Their relative contribution as a function of $s$ and $t$ has been quantified in Sec.~\ref{pom_dom} Figs.~\ref{fig:Rt0}.

4. To summarize, our predictions for the LHC are: 
\small{
\begin{table}[tbph!]
\begin{center}
\begin{tabular}{|c|c|c|c|c|c|}
  \hline
  & $\sigma_{tot}$ (mb) & $\sigma_{el}$ (mb) & $\sigma_{inel}$ (mb) & $\sigma_{el}\over\sigma_{tot}$&$\rho$\\
  \hline
  7 TeV  & $98\pm1$  & $26$ & $72$ & $0.27$ & $0.16$\\
  14 TeV & $111\pm2$ & $32$ & $79$ & $0.29$ & $0.16$\\
  \hline
\end{tabular}\\
\vspace{0.5cm}
\begin{tabular}{|c|c|c|c|}
  \hline
  & $B\left(t=0.1\right)$ GeV$^{-2}$ & $B\left(t=0.3\right)$ GeV$^{-2}$ & $-t_{min}$ GeV$^2$ \\
  \hline
  7 TeV  & $19.2$ & $19.6$ & $0.65$ \\
  14 TeV & $20.4$ & $20.9$ & $0.60$ \\
  \hline
\end{tabular}\\

\end{center}
\caption{Predictions for total elastic, inelastic $pp$ cross sections, local slope and the position of the diffractive minimum calculated in model with the parameters presented in Table (\ref{tab:fitParam1}).}
\label{tab:LHCpredictions}
\end{table}
}

\section*{Acknowledgements}
We thank Tam\'as Cs\"{o}rg\H{o} and Mario Deile for useful discussions and correspondence. Also we thank Jan Ka\v{s}par for Monte Carlo implementation of the model. L.J. thanks also Karsten Eggert for his critical remarks. The work of L. Jenkovszky was supported also by the Hungarian Science Abroad Project.

\newpage
\small{
\begin{table}[tbph!]
\begin{center}
\begin{tabular}{|c|c|c|c|c|c|}
  \hline
  \multicolumn{2}{|c|}{Pomeron} & \multicolumn{2}{|c|}{Odderon} & \multicolumn{2}{|c|}{Reggeons} \\
  \hline
  $a_P$ & $262$ & $a_O$ & $0.088$ & $-a_f$ & $12.6$ \\
  $b_P$ [GeV$^{-2}$] & $8.4$ & $b_O$ [GeV$^{-2}$] & $14.2$ & $b_f$ [GeV$^{-2}$] & $4.4$          \\
  $\delta_P$ & $0.05$ & $\delta_O$ & $0.17$ & $ - $ & $ - $ \\
  $\alpha_{1P}$ & $0.44$ & $\alpha_{1O}$& $0.043$ & $a_{\omega}$ & $8.2$ \\
  $\varepsilon_P$ & $0.015$ & $\varepsilon_O$ & $0.$ & $ b_\omega $ [GeV$^{-2}$] & $23.8$         \\
  $s_P$ [GeV$^2$] & $100$ & $s_{O}$ [GeV$^2$] & $100$& $s_0$ [GeV$^2$] & $1$\\
  \hline
  \hline
  & $\sigma_{tot},\sigma_{\rho},\frac{d\sigma_{pp}}{dt},\frac{d\sigma_{p\bar p}}{dt}$ &  &  &  & \\
  \hline
  $\chi^2/NDF$ & $3.2$ &  &  &  &  \\
  \hline
\end{tabular}\\
\end{center}
\caption{Fitted parameters of the model with trajectory \ref{eq:tr1}.}
\label{tab:fitParam1}
\end{table}
}

\small{
\begin{table}[tbph!]
\begin{center}
\begin{tabular}{|c|c|c|c|c|c|}
  \hline
  \multicolumn{2}{|c|}{Pomeron} & \multicolumn{2}{|c|}{Odderon} & \multicolumn{2}{|c|}{Reggeons} \\
  \hline
  $a_P$ & $253$ & $a_O$ & $0.11$ & $-a_f$ & $12.4$ \\
  $b_P$ [GeV$^{-2}$] & $8.4$ & $b_O$ [GeV$^{-2}$] & $14$ & $b_f$ [GeV$^{-2}$] & $4.0$          \\
  $\delta_P$ & $0.05$ & $\delta_O$ & $0.16$ & $ - $ & $ - $ \\
  $\alpha_{1P}$ & $0.41$ & $\alpha_{1O}$& $0.046$ & $a_{\omega}$ & $8.0$ \\
  $\alpha_{2P}$ [GeV$^{-1}$]  & $3.34$ & $\alpha_{2O}$ [GeV$^{-2}$] & $-$ & $b_{\omega}$ [GeV$^{-2}$] & $15.4$ \\
  $\alpha_{3P}$ [GeV$^{2}$] & $0.14$ & $-$ & $-$ & $-$ & $-$ \\
  $\varepsilon_P$ & $0.017$ & $\varepsilon_O$ & $-$ & $ - $ & $-$         \\
  $s_P$ [GeV$^2$] & $100$ & $s_{O}$ [GeV$^2$] & $100$& $s_0$ [GeV$^2$] & $1$\\
  \hline
  \hline
  & $\sigma_{tot},\sigma_{\rho},\frac{d\sigma_{pp}}{dt},\frac{d\sigma_{p\bar p}}{dt}$ &  &  &  & \\
  \hline
  $\chi^2/NDF$ & $3.1$ &  &  &  &  \\
  \hline
\end{tabular}\\
\end{center}
\caption{Fitted parameters of the model with trajectory \ref{eq:tr2}.}
\label{tab:fitParam2}
\end{table}
}
\small{
\begin{table}[tbph!]
\begin{center}
\begin{tabular}{|c|c|c|c|c|c|}
  \hline
  \multicolumn{2}{|c|}{Pomeron} & \multicolumn{2}{|c|}{Odderon} & \multicolumn{2}{|c|}{Reggeons} \\
  \hline
  $a_P$ & $258$ & $a_O$ & $0.0386$ & $-a_f$ & $12.4$ \\
  $b_P$ [GeV$^{-2}$] & $8.6$ & $b_O$ [GeV$^{-2}$] & $20.8$ & $b_f$ [GeV$^{-2}$] & $4.3$          \\
  $\delta_P$ & $0.05$ & $\delta_O$ & $0.16$ & $ - $ & $ - $ \\
  $\alpha_{1P}$ & $1.33\cdot10^{4}$ & $\alpha_{1O}$& $8.86\cdot10^{3}$ & $a_{\omega}$ & $8.1$ \\
  $\alpha_{2P}$ [GeV$^{-2}$] & $3.2\cdot10^{-5}$& $\alpha_{2O}$ [GeV$^{-2}$] & $3.66\cdot10^{-6}$ & $b_{\omega}$ [GeV$^{-2}$] & $282.3$ \\
  $\varepsilon_P$ & $0.02$ & $\varepsilon_O$ & $0.47$ & $ - $ & $-$         \\
  $s_P$ [GeV$^2$] & $100$ & $s_{O}$ [GeV$^2$] & $100$& $s_0$ [GeV$^2$] & $1$\\
  \hline
  \hline
  & $\sigma_{tot},\sigma_{\rho},\frac{d\sigma_{pp}}{dt},\frac{d\sigma_{p\bar p}}{dt}$ &  &  &  & \\
  \hline
  $\chi^2/NDF$ & $3.08$ &  &  &   &  \\
  \hline
\end{tabular}\\
\end{center}
\caption{Fitted parameters of the model with trajectory \ref{eq:tr3}.}
\label{tab:fitParam3}
\end{table}
}

\vfill \eject

\newpage
\begin{thebibliography}{99}

\bibitem{review} R.~Fiore {\it et al.}
Int. J. Mod. Phys., A24 (2009) 2551-2559, arXiv:hep-ph/0812.0539.

\bibitem{DL} S.~Donnachie and P.~Landshoff, Phys. Lett. {\bf B 123} (1983) 345; Nucl. Phys. {\bf 267} (1985) 690.

\bibitem{KKL} J.~Kontros, K.~Kontros, and A.~Lengyel, {\it Model of realistic Reggeons for $t=0$ high-energy scattering}, hep-ph/0104133.

\bibitem{C-Y} T.T.~Chou and C.N.~Yang, Phys .Rev.Lett. {\bf 20} (1968) 1615.

\bibitem{PB} R.J.J.~Phillips and V.~Barger, Phys. Lett. {\bf 46B}
(1973) 412.

\bibitem{KKL1} J.~Kontros, K.~Kontros, and A.~Lengyel, {\it Pomeron model and exchange degeneracy of $R$ trajectories,} hep-ph/0006141.

\bibitem{J-Vall} L.L.~Jenkovszky and A.N.~Vall, ITP Preprint, Kiev 1974;
Czech. J. Phys. B {\bf 26} (1976) 447.

\bibitem{reviews} L.~Jenkovszky, Fortschritte der Physik, {\bf 34} (1986);\\
L.~Jenkovszky, Rivista Nuovo Cim. {\bf 10} (1987) 1;\\
L.~Jenkvoszky, EChAYa (Egl. translation: PEPAN) {\bf 34} (2003) 1196.

\bibitem{VJS} A.N.~Vall, L.L.~Jenkovszky and B.V.~Struminsky, EChAYa (Russian translation: PEPAN) {\bf 19} (1988) 180.

\bibitem{KLT} K.~Kontros, A.~Lengyel and Z.~Tarics, {\it $pp$ and $p\bar p$ elasctic scattering in a multipole Pomeron model}, hep-ph/0011398.

\bibitem{2} P.~Desgrolard, M.~Giffon, L.L.~Jenkovszky, Z. Phys. C
{\bf 55}(1992) 643.

\bibitem{3} R.J.M.~Covolan, P.~Desgrolard, M.~Giffon, L.L.~Jenkovszky and E.~Predazzi,
Z.Phys. C {\bf 58} (1993) 109.


\bibitem{Cohen} G.~Cohen-Tannoudji {\it et al.} Lettere Nuovo Cim. {\bf 5} (1972) 957.

\bibitem{Zurab} A.I.~Bugrij, L.L.~Jenkovszky and Z.E.~Chikovani, Z.Phys.
C, Particles and Fields, {\bf 4} (1980) 45.

\bibitem{DLM} P.~Desgrolard, A.I.~Lengyel and E.S.~Martynov, Nuovo Cim. A {\bf 110} (1997) 251.


\bibitem{DJS} P.~Desgrolard, L.L.~Jenkovszky and B.V.~Struminsky, Z. Phys. C {\bf 11} (1999) 145.


\bibitem{Amaldi:1979kd} U.~Amaldi and K.R.~Schubert,  Nucl.\ Phys.\  B {\bf 166} (1980) 301.

\bibitem{Albrow:1976sv} M.G.~Albrow {\it et al.},  Nucl.\ Phys.\  B {\bf 108} (1976) 1.

\bibitem{Breakstone:1984te} A.~Breakstone {\it et al.}, Nucl.\ Phys.\  B {\bf 248} (1984) 253.

\bibitem{Breakstone:1985pe} A.~Breakstone {\it et al.}, Phys.\ Rev.\ Lett.\  {\bf 54} (1985) 2180.

\bibitem{Bozzo:1985th} M.~Bozzo {\it et al.}, Phys.\ Lett.\  B {\bf 155} (1985) 197.

\bibitem{Bernard:1986ye} D.1.~Bernard {\it et al.},  Phys.\ Lett.\  B {\bf 171} (1986) 142.

\bibitem{Amos:1990fw} N.A.~Amos {\it et al.}, Phys.\ Lett.\  B {\bf 247} (1990) 127.

\bibitem{Abe:1993xx} F.~Abe {\it et al.}, Phys.\ Rev.\  D {\bf 50} (1994) 5518.

\bibitem{data} http://qcd.theo.phys.ulg.ac.be/$\sim$cudell/;\\
http://qcd.theo.phys.ulg.ac.be/$\sim$cudell/DATA.html/;\\
http://www.theo.phys.ulg.ac.be/$\sim$cudell/data/;\\
http://pdg.lbl.gov/2002/.

\bibitem{MINUIT} F.~James and M.~Roos, Comput.\ Phys.\ Commun.\  {\bf 10}, (1975) 343-367;

\bibitem{TT} S.M.~Troshin and N.E.~Tyurin, {\it The role of double helicity-flip amplitudes in small-angle elastic $pp$-scattering}, Phys.\ Lett.\  B {\bf 580} (2004), 54, hep-ph/0310086\\
J.R.  Cudell, E.  Predazzi and O.V.  Selyugin, {\it The high-energy hadron
spin-flip amplitude at small momentum transfer and new AN data from RHIC}, Eur. Phys. J. A {\bf 21} (2004) 479, hep-ph/0401040;\\
J.R. Cudell, E. Predazzi and O.V. Selyugin, {\it Interactions at large distances and spin effects in nucleon-nucleon and nucleon-nuclei scattering}, Phys. Part. Nucl. {\bf 35} (2004) 75, hep-ph/0312195.

\bibitem{kontros} L.~Fajardo {\it et al.}, Phys.\ Rev.\  D {\bf 24} (1981) 46;\\
J.E.~Kontros and A.I.~Lengyel, in ``Strong interactions at long distances'', L.L.~Jenkovszky editor, Hadronic Press, Palm Harbor (1995), p.67; Ukr. J. Phys., v.41 (1996).

\bibitem{Ilyin} M.~Bertini, P.~Desgrolard and Yu.M.~Ilyin, International J. Phys. v.1 (1 dec. 1995) 45.

\bibitem{ATLAS} ATLAS Collab. {\it Measurement of the Inelastic Proton-Proton Cross-Section at $\sqrt{s}$ = 7 TeV with the
ATLAS Detector}, hep-ph/1104.0326.

\bibitem{CMS} CMS at DIS11 https://wiki.bnl.gov/conferences/images/8/8c/Parallel.SxDNVM.Marone.13April.talk.pdf

\end{thebibliography}
\end{document}